# Design of Switchable Frequency-Selective Rasorber with A-R-A-T or A-T-A-R Operating Modes

Xiangkun Kong, *Member*, *IEEE*, Xin Jin, Xuemeng Wang, Weihao Lin, He Wang, Zuwei Cao and Steven Gao, *Fellow, IEEE*

*Abstract*—**This work presents a switchable frequency-selective rasorber (SFSR) with two operating modes**. **Switching from transmission to reflection can be achieved by appropriately adding feeders and PIN diodes based on cascaded two-dimensional lossy and lossless frequency-selective surface (FSS) screens. The proposed SFSR** can realize out-of-band absorption. Analysis of the equivalent circuit model (ECM) can be useful for achieving a switchable rasorber. As the state of the PIN diodes changes, the working state of the SFSR can be switched from low-frequency reflection and high-frequency transmission to low-frequency transmission and high-frequency reflection, respectively. **In the final-revision simulation of the working band of the SFSR, in the ON state, reflection and transmission peaks of -0.55 dB at 4.06 GHz and -0.52 dB at 5.97 GHz, respectively, are achieved; in the OFF state, the transmission and reflection peaks of -0.31 dB at 4.04 GHz and -0.26 dB at 6.01 GHz, respectively, are obtained. A prototype sample is developed and validated. The results are in good agreement with those of the full-wave simulation. The proposed design can be used in intelligent anti-jamming communication.**

## I. INTRODUCTION

FREQUENCY-selective rasorbers (FSRs) can be designed as radomes with low insertion within the absorption band, which can reduce the radar cross-section (RCS) in a radar antenna system. According to the location of the transmission response in the absorption band, FSRs can be classified into three categories: transmission below the absorption band (T-A) [1][2], transmission between two absorption bands (A-T-A) [3]–[9], and transmission above the absorption band (A-T) [10][11].

With the development of shared-surface dual-band antenna systems for 5G applications, the demand for dual-band or multi-band FSRs has increased because of their compact size, switchability, and easy integration [12]. However, these single-band FSRs are unsuitable for multiband radar antenna systems. In recent years, dual-band FSRs have attracted considerable attention [13]–[17]. It is well known that lumped components, including inductance and capacitance (LC), can be utilized to realize series resonance and parallel resonance. Based on the above principle, capacitances are connected parallel to the metal wire, leading to the generation of two transmission bands [13]. Owing to the use of capacitance instead of metal parallel resonance, a miniaturized dual-band FSR is realized. However, the measured passbands are prone to frequency shifts owing to the deviation in the capacitance value. In addition, the parallel resonances of the metal structure are used to generate two transmission bands [14]–[16]. In [17], these two methods have been combined to generate dual transmission bands via the capacitance and metal gap. The parallel metal patch in this configuration is analogous to the capacitance in the equivalent circuit model. Consequently, two transmission bands are produced by the three series-connected LCs.

Although the dual-band FSRs described above can be designed as passive radomes, their application is limited to multifunctional systems. Generally, PIN diodes are chosen as switches in the FSR [18]–[20]. The design of a switchable frequency-selective rasorber (SFSR) with A-R-A-T or A-T-A-R operating mode was proposed in this study to realize a reconfigurable wave-transmission window and strong reflection band in the wide -10 dB reflection band of the SFSR. The novelty of this work can be summarized as follows: 1) Switching diodes are applied to control the transmission window and reflection band of the rasorber. In this case, the proposed switchable structure can be operated in two operating modes based on the engineering requirements. 2) Unlike the previously designed typically lossy layers in switchable FSRs, the proposed design has two wave-transmitting windows corresponding to different operating principles. 3) Fewer lumped elements are required in one unit, which simplifies the complexity of the SFSR. The manufactured prototype was tested and validated through simulations. The designed SFSR satisfies the engineering requirements of an intelligent anti-jamming system [21][22]. A reconfigurable wave-transmission window and a strong reflection band in the wide -10 dB reflection band of the SFSR are achieved. The proposed design can realize intelligent anti-jamming communication and shield high-power electromagnetic interference attacks.

This work was supported in part by National Natural Science Foundation of China under Grant 62071227, in part by National Science Foundation of Jiangsu Province of China under Grant BK20201289, in part by the Postgraduate Research & Practice Innovation Program of Jiangsu Province under Grant SJCX20_0070 and in part by the Fundamental Research Funds for the Central Universities under Grant kfjj20200403.

X. Kong, X. Jin, X. Wang, W. Lin, H. Wang, Z. Cao are with the College of Electronic and Information Engineering, Nanjing University of Aeronautics and Astronautics, Nanjing, 211106, China (e-mail: xkkong@nuaa.edu.cn).
S. Gao is with the University of Kent, Canterbury CT2 7NT, U. K. (e-mail: s.gao@kent.ac.uk).

## II.  DESIGN OF THE SFSR

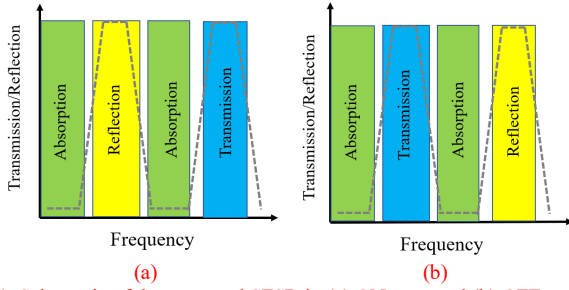

(a)                                              (b)

Fig. 1. Schematic of the proposed SFSR in (a) ON state and (b) OFF state.

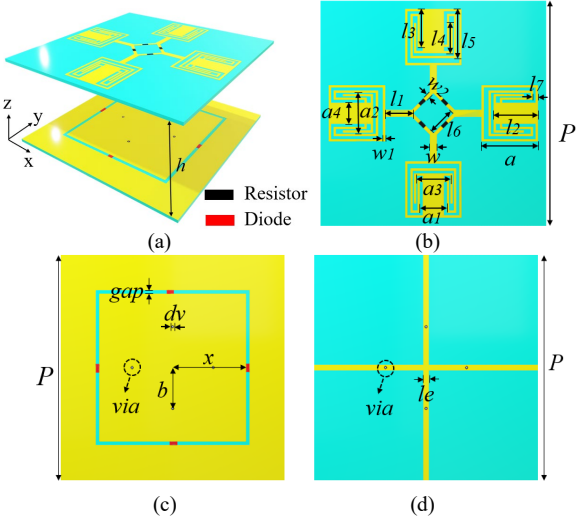

Fig. 2. Proposed SFSR. (a) 3D view. (b) Top view of the lossy layer. (c) The top view and (d) the bottom view of the switchable frequency-selective surface.

A schematic of the transmission/reflection responses of the SFSR is shown in Fig. 1. When the PIN diodes are in the OFF state, low-frequency transmission at approximately 4 GHz and high-frequency reflection around 5.6 GHz were acquired in the A-T-A-R mode. In contrast, low-frequency reflection and high-frequency transmission were acquired in the A-R-A-T mode when the PIN diodes are in the ON mode.

A dual-band lossy layer was designed based on the resonance principle [14]-[16]. The SFSR is composed of a dual-band lossy layer and a switchable FSS separated by an air gap with a thickness of $h$=15 mm, as shown in Fig. 2. Copper (0.035 mm) of the top and bottom layers was printed on the F4BM substrate ($\varepsilon_r$ = 2.2, $tan\,\delta$ = 0.001, $thickness$=0.5 mm). The lossy layer was constructed using four resistors ($R$=300 Ω) to realize the absorption bands, as shown in Fig. 2(b). The diodes were welded to the metal crevices of the FSS. In Fig. 2(c), the diodes SMP1320-079LF (ideal simplified equivalent parameters: $L_{chip}$ = 0.7 nH, $C_{off}$ = 0.23 pF, $R_s$ = 0.75 Ω [23]) produced by Skyworks were selected as the switch. The four black circles shown in Fig. 2(c) and (d) are vias that connect the bottom feedline to the supply voltage of the diodes. By supplying different voltages to the positive and negative poles of the diodes, the resonant frequency of the switchable FSS is transformed from one band to another. The dimensions are as follows: $P$= 32 mm, $w$= 1 mm, $w_1$=0.33 mm, $w_2$= 0.5 mm, $a$= 8 mm, $a_1$= 3.83 mm, $a_2$= 5.83 mm, $a_3$= 4.83 mm, $a_4$= 3 mm, $l_1$=

4 mm, $l_2$= 6.3 mm, $l_3$= 5.4 mm, $l_4$= 4.3 mm, $l_5$= 7 mm, $l_6$= 4 mm, $l_7$= 0.7 mm, $x$= 10.5 mm, $b$= 5.8 mm, $dv$= 0.4 mm, $le$= 0.8 mm, $gap$=0.5mm. The simulations of a linearly polarized normal incidence propagating toward the -z direction are presented in Fig. 3. Indeed, owing to the structural symmetry, the simulated results for the TE and TM polarizations are nearly identical in the ON and OFF states. Here, only the TE polarization results are provided and discussed. The simulations are illustrated in Fig. 3(a), when a forward bias is applied to the diodes. The reflection and transmission peaks in the ON state are -0.73 dB at 4.03 GHz and -0.41 dB at 5.63 GHz. With a -10 dB reduction, the absorption bands are 2.79 GHz to 3.46 GHz and 4.47 GHz to 5.01 GHz. The simulations are in Fig. 3(b) when a reverse bias is applied to the diodes. The transmission and reflection bands have peaks of -0.33 dB at 4.03 GHz and -0.39 dB at 5.64 GHz, respectively. The absorption bands range from 2.58 GHz to 3.52 GHz and 4.56 GHz to 5.19 GHz with -10 dB reduction.

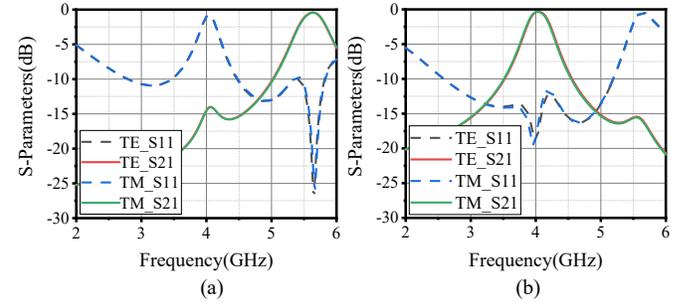

Fig. 3. The simulated results of the SFSR in (a) ON state and (b) OFF state.

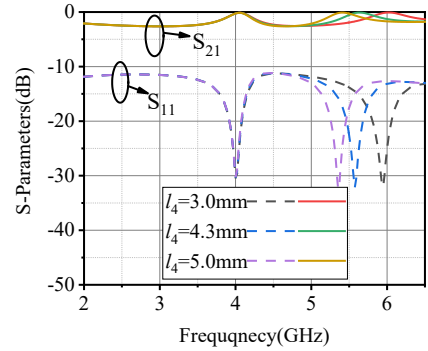

Fig. 4. The simulations of the lossy layer with respect to different values $l_4$.

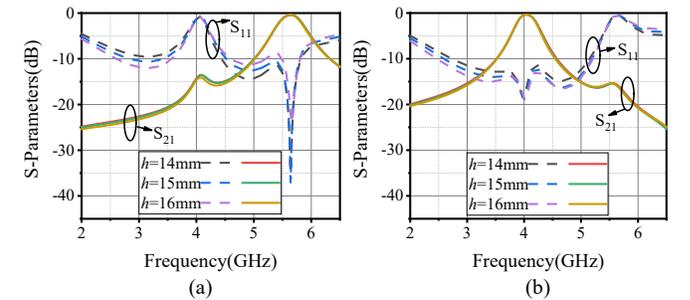

Fig. 5. Simulations of the air layer with respect to different values $h$. (a) ON state. (b) OFF state.

The dimensions $l_4$ of the lossy layer were investigated, which resulted in a shift in the upper transmission frequency, as illustrated in Fig. 4. As the value of $l_4$ increased, the upper transmission shift to a lower frequency. Meanwhile, the lower transmission response is fairly constant. The effects of $h$ are illustrated in Fig. 5. With an increase in the thickness of the air layer, the lower absorption bandwidth became wider.

## III. DESIGN AND ANALYSIS OF EQUIVALENT CIRCUIT

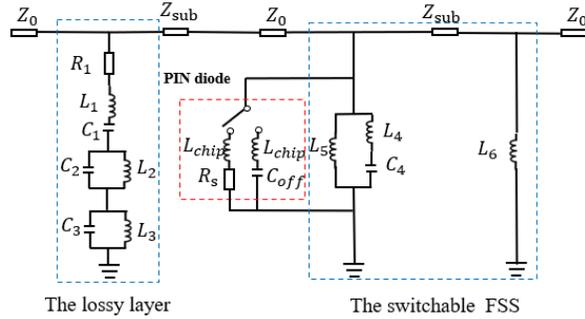

Fig. 6. ECM of the SFSR. (optimized circuit parameters: $R_1$=445.6 $\Omega$, $L_1$=16.219 nH, $C_1$=0.1 pF, $L_2$=2.03 nH, $C_2$=0.73 pF, $L_3$=0.655 nH, $C_3$=1.2 pF, $L_4$=0.3 nH, $C_4$=1.15 pF, $L_5$=0.9 nH, $L_6$=24.076 nH, $Z_{sub}$=377/$\sqrt{2.2}$ $\Omega$, $Z_0$=377$\Omega$.)

An equivalent circuit model was established to explain the design better. LC parallel resonances and LC series resonance with resistance are extensively utilized to realize a lossy layer of the SFSR. In ECM, parallel and series resonances produce transmission bands and absorption, respectively. Resultantly, two parallel resonances were used to provide dual transmission bands. Figure 6 depicts the final ECM concept. In the ON state, the simplified diode models are equivalent to $L_{chip}$ and $R_s$ in series, whereas in the OFF state is equivalent to $L_{chip}$ and $C_{off}$ in series.

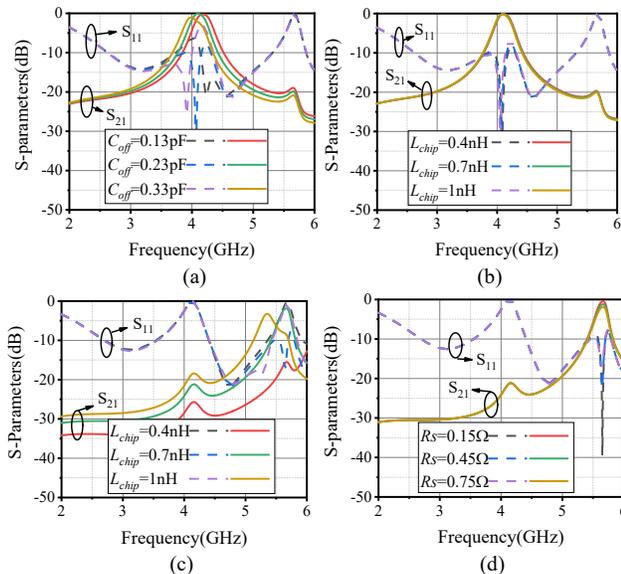

Fig. 7. Influence of the different diodes' parasitic values in ECM. In OFF state, (a) $L_{chip}$=0.7 nH and variable $C_{off}$; (b) $C_{off}$=0.23 pF and variable $L_{chip}$. In ON state, (c) $R_s$=0.75 $\Omega$ and variable $L_{chip}$, (d) $L_{chip}$=0.7 nH and variable $R_s$.

The parasitic elements of the purchased diode are unclear. Therefore, the values of $L_{chip}$, $R_s$, and $C_{off}$ that cause changes in the results are investigated in the ECM, as shown in Fig. 7. Within a certain range, when the diode in the OFF state and $L_{chip}$ are fixed as a constant, variable $C_{off}$ is the most crucial element affecting the low transmission band, whereas variable $L_{chip}$ has minimal effect when $C_{off}$ is constant, as depicted in Fig. 7(a) and

(b). Similarly, in the ON state, the variable $L_{chip}$ modifies the upper transmission band, and the variable $R_s$ primarily causes high-frequency insertion loss, as illustrated in Fig. 7(c) and (d).

## IV. FABRICATION AND EXPERIMENTAL VERIFICATION

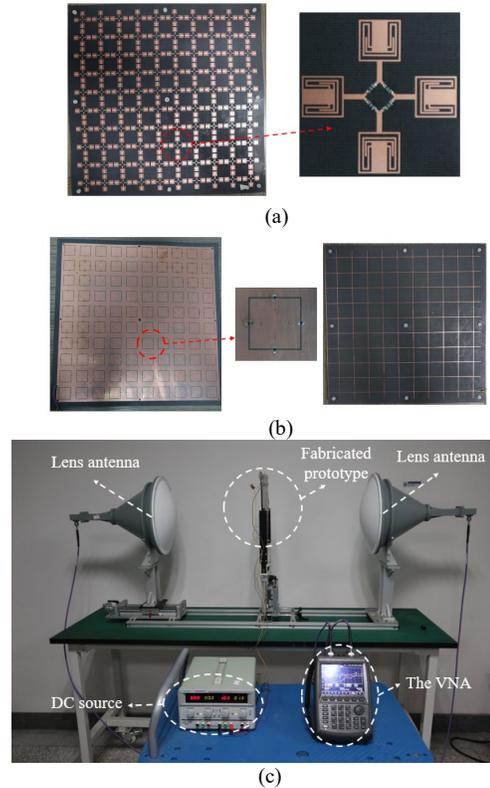

Fig. 8. Fabricated prototype and measurement. (a) Two transmission bands' lossy layer. (b) Both side views of the switchable FSS. (c) Measurement environment.

The performance of the SFSR was validated by manufacturing a prototype with dimensions of 340 mm × 340 mm, including 10 × 10 unit cells, as shown in Fig. 8. The prototypes of the lossy and switchable layers are depicted in Fig. 8 (a) and (b), respectively. The gap between the lossy and switchable layers was secured using nine plastic screws. The lumped resistors were soldered between the metal gaps of the lossy layer. The SMP1320-079LF diodes were soldered onto the switchable FSS. A DC source was employed to control the states of the PIN diodes to operate them properly. The prototype was measured using two pairs of lens antennas (2~4 and 4~8 GHz), as shown in Fig. 8(c). The configuration and measurement of the lens antenna system are described in detail in [24][25].

From the analysis of the ECM in Section III, the parasitic parameter values of the diodes must be considered; otherwise, frequency deviation will damage the final results. Thus, the switchable FSS is primarily manufactured and measured to ensure transmission response in various diode states. Figure 9 shows the simulations and measurements. Unpredictable parasitic parameters cause discrepancies between simulations and measurements. The measured transmission band of the switchable FSS is at 5.96 GHz in ON state, existing deviation, as shown in Fig. 9(a). The measured transmission band in the

TABLE I
COMPARISON WITH OTHER STRUCTURES

| Ref | S/T | Number of the switchable band | Operating bandwidth | Frequency Response | Max. RT(dB) | Polarization insensitive | Thickness (@$f_L$) | Number of Lumped elements in one unit |
|---|---|---|---|---|---|---|---|---|
| [13] | No | 0 | 123% | A-T-A-T-A | -1.2/-0.82 | Yes | 0.077$\lambda_L$ | 12 |
| [14] | No | 0 | 122.2% | A-T-A-T-A | -0.15/-0.31 | Yes | 0.091$\lambda_L$ | 4 |
| [15] | No | 0 | 117.7% | A-T-A-T-A | -0.25/-0.1 | Yes | 0.076$\lambda_L$ | 4 |
| [16] | No | 0 | 87.6% | A-T-A-T | -0.39/-0.64 | Yes | 0.117$\lambda_L$ | 4 |
| [17] | No | 0 | 100% | A-T-A-T | -0.33/-0.92 | No | 0.099$\lambda_L$ | 3 |
| [18] | Yes | 1 | ON:120% OFF:120% | A-T/R | ON:-0.34 OFF:N.A. | Yes | 0.158$\lambda_L$ 0.158$\lambda_L$ | 8 |
| [19] | Yes | 1 | ON:92.6% OFF:90.14% | A-T/A-A | ON:N.A. OFF:-0.45 | Yes | 0.080$\lambda_L$ 0.082$\lambda_L$ | 12 |
| [20] | Yes | 2 | ON:62.2% OFF:76.2% | A-T/R-A | ON:N.A. OFF:-0.62 | Yes | 0.168$\lambda_L$ 0.141$\lambda_L$ | 16 |
| **This work** | **Yes** | **2** | **ON:70.6% OFF:80%** | **ON:A-R-A-T OFF:A-T-A-R** | **-0.55/-0.52 -0.31/-0.26** | **Yes** | **0.153$\lambda_L$ 0.137$\lambda_L$** | **8** |

S/T: switchable or tunable; A: absorption; T: transmission; R: reflection; RT: simulated reflection or transmission coefficient; $f_L$ is the lowest frequency.

OFF state is almost identical to that in the simulation at 4.04 GHz, as shown in Fig. 9(b). However, because the dielectric substrate is just 0.5 mm thick, the bent substrate cannot maintain the switchable FSS parallel to the calibration surface. Consequently, the measured bands outside the transmission band were worse than the corresponding bands in the simulations.

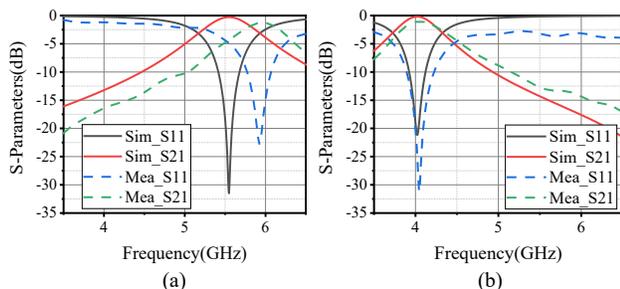

Fig. 9. The measurements and simulations of the switchable FSS with (a) diodes ON and (b) diodes OFF.

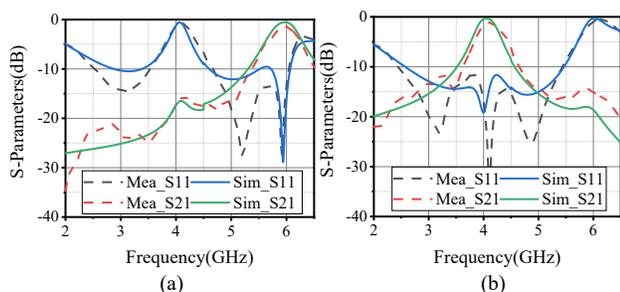

Fig. 10. Measured and simulated results of the SFSR with (a) diodes ON and (b) diodes OFF.

Compared with the simulation, the measured passband of the switchable FSS shifted in the ON state and changed slightly in the OFF state. Revising the simulation model is necessary, and there are two steps: matching the simulation of the switchable FSS with measured results ($x$=9.75 mm in ON state, $x$=10.5 mm in the OFF state). Subsequently, the two passbands of the lossy layer were revised following the revised switchable FSS ($l_4$= 3 mm). The final-revision segmented simulations of the SFSR are shown in Fig. 10. When the PIN diodes are in the ON state, the reflection and transmission peaks of the operating band are at 4.06 GHz (-0.55 dB) and 5.97 GHz (-0.52 dB), respectively. The absorption bands are from 2.87 GHz to 3.38 GHz and 4.56 GHz to 5.3 GHz with a -10 dB reduction. The simulations corresponding to a case in which a reverse bias is applied to the diodes are depicted in Fig. 10(b). The transmission and reflection bands have peaks of -0.31 dB at 4.04 GHz and -0.26 dB at 6.01 GHz, respectively. With a -10 dB reduction, the absorption bands are at 2.57 GHz to 3.51 GHz and 4.62 GHz to 5.46 GHz, respectively.

In experiment, 1.3 V forward bias voltage and 10 V reverse bias voltage were applied on the diodes. The simulated and measured findings were compared and found to be in high agreement, as shown in Fig. 10. The measured absorption bands in the ON state are 2.53 GHz to 3.53 GHz and 4.67 GHz to 5.36 GHz. The reflection and transmission bands have peaks of -0.75 dB at 4.1 GHz and -1.29 dB at 5.95 GHz, respectively. The observed absorption bands in the OFF state are at 2.43 GHz to 3.62 GHz and 4.65 GHz to 5.43 GHz. The transmission and reflection bands have peaks of -1.15 dB at 4.09 GHz and -0.46 dB at 6.11 GHz, respectively. The simulated low absorption band was worse than the measured one because of the uneven air layer, as discussed in Fig. 5. TABLE I compares this work with previous works. The proposed SFSR, including the A-R-A-T and A-T-A-R modes, is reconfigurable and polarization-insensitive.

## V. CONCLUSION

In this study, an SFSR including the A-R-A-T or A-T-A-R mode was proposed. Metal resonances were employed to realize the two transmission bands of the lossy layer. The effect of the parasitic elements of the diodes on the transmission response in ECM was investigated. A prototype based on the ECM was fabricated. A distinct production approach was used to minimize the frequency deviation caused by unknown causes. A prototype of the proposed SFSR was developed and tested. The measurement results the and revised simulation results were in agreement.